# Symmetry-breaking effects on spin-orbit torque switching in ferromagnetic semiconductors with perpendicular magnetic anisotropy

Apu Kumar Jana and Sanghoon Lee[*]

*Physics Department, Korea University, Seoul 136-701, Republic of Korea.*

**Abstract:**

This study explores the mechanisms of spin-orbit torque (SOT) switching in ferromagnetic semiconductors (FMS) with perpendicular magnetic anisotropy (PMA), emphasizing the impact of symmetry-breaking. Using micromagnetic simulations based on the Landau-Lifshitz-Gilbert (LLG) equation, we examine several symmetry-breaking factors, including bias field misalignment, interlayer exchange coupling, out-of-plane spin polarization, and tilted magnetic anisotropy. The results reveal that bias field misalignment relative to the film plane significantly distorts the SOT switching hysteresis. Additionally, intrinsic symmetry-breaking effects, such as internal coupling fields, out-of-plane spin polarization, and tilted anisotropy, facilitate field-free SOT (FF-SOT) switching without external bias fields. Each type of FF-SOT switching exhibits distinct characteristics, including hysteresis shifts, switching ratios, and saturated magnetization. This work emphasizes the role of symmetry-breaking in FF-SOT switching and offers fundamental information for interpreting FF-SOT switching observed from FMS films in experiments, contributing to the optimization of SOT efficiency and the advancement of spintronics technologies.

[*] Author to whom correspondence should be addressed: slee3@korea.ac.kr

**Introduction**

Spin-orbit torque (SOT) is an advanced concept for manipulating magnetization using an electric current,[1–5] offering significant advantages over conventional methods. SOT enables faster write speeds, higher endurance, and greater energy efficiency, making it ideal for potential applications like SOT magneto-resistive random-access memory (SOT-MRAM), where the write and read paths are separated.[6–9] This SOT-driven magnetization manipulation also holds great promise for transforming other spintronics devices, such as nano-oscillators, memristors, and THz generators.[10–12] While SOT enables magnetization control for both perpendicular and in-plane magnetic anisotropy systems, perpendicular systems are particularly appealing for SOT-MRAM applications due to their speed and scalability.[12] However, achieving SOT-based switching in perpendicular magnetization requires breaking the symmetry between the up and down magnetization states,[13] which is typically done by applying an in-plane magnetic bias field—a step that is undesirable for practical applications.[14]

Many approaches have been proposed to achieve SOT magnetization switching without the need for an external bias field, and these methods can be classified into several main categories: (a) utilizing a built-in in-plane (IP) field through coupling effects, (b) inducing lateral structural asymmetry, (c) utilizing the out-of-plane component of spin polarization, and (d) forming hybrid structures.[1] The built-in coupling field can be generated by introducing either a ferromagnetic layer with in-plane anisotropy (IPA) or antiferromagnetic layers, leading to exchange coupling at the interface between two magnetic layers.[15,16] For lateral symmetry breaking, a wedge-shaped layer can be introduced, which induces either an out-of-plane effective field or tilted anisotropy.[17] The out-of-plane component of spin polarization can be achieved in ferromagnetic systems with an inversion asymmetric crystal structure, such as $WTe_2$, $IrMn_3$, $Mn_2Au$.[18–20] Finally, hybrid approaches combine ferroelectric/ferromagnetic structures or exploit combined spin-transfer torque (STT) and SOT effects.[21]

While field-free spin-orbit torque (FF-SOT) magnetization switching in ferromagnetic (FM)/heavy metal (HM) bilayer systems is well-established in both experiments [4,11,16,18,22] and theory,[14,23,24] FF-SOT switching in ferromagnetic semiconductor (FMS) systems—a newly emerging material platform for SOT research—has only recently been observed by two research groups[25,26] and is not yet fully understood. The complex mechanisms underlying carrier spin polarization and SOT phenomena in FMS systems are still under debate, with limited clarity on the origin of FF-SOT switching in these FMS films. For instance, FF-SOT switching observed in GaAs-based FMS films with perpendicular magnetic anisotropy (PMA) by M. Jiang et al.[26] attributed to a combination of Dzyaloshinskii-Moriya interaction (DMI) [27] and tilted anisotropy, whereas Lee et al.[25] proposed that the observed FF-SOT switching in (Ga,Mn)(As,P) film with PMA was driven by an internal coupling field, potentially arising from unexpected Mn-oxide magnetic structures formed on the surface of the film during annealing. In addition, FMS films grown on vicinal crystal surfaces[28] can also exhibit FF-SOT switching due to tilted anisotropy or an additional out-of-plane component of spin polarization. Thus, identifying the exact origin of FF-SOT switching in FMS films remains a challenge due to complex symmetry-breaking effects.

This study addresses these challenges through micromagnetic simulations, analyzing various symmetry-breaking mechanisms, including bias field misalignment, exchange coupling, out-of-plane spin polarization, and tilted magnetic anisotropy. The results provide crucial insights into the origins of FF-SOT switching in FMS films and contribute to a deeper understanding of the underlying physics, offering guidance for future experimental interpretations and advancements in spintronics technologies.

**Simulation Methods**

To investigate the SOT switching mechanism in FMS film, we simulated the current scan hysteresis loops by solving the Landau-Lifshitz-Gilbert (LLG) equation with additional effects such as bias field, out-of-plane spin polarization, and tilted magnetic anisotropy. The LLG equation, incorporating SOT contributions, is expressed as[13]

$$\frac{d\hat{m}}{dt} = -\frac{\gamma}{1+\alpha^2}[\hat{m} \times \hat{H} + \alpha\hat{m} \times \hat{m} \times \hat{H} + (1-r)\zeta J(\hat{m} \times \hat{\sigma}) + r\zeta J(\hat{m} \times \hat{m} \times \hat{\sigma})] \ldots\ldots (1),$$

where the 1$^{st}$ term is the precession term, 2$^{nd}$ term is the damping term, 3$^{rd}$ term is the field-like torque term (FLT) and 4$^{th}$ term is the damping-like torque (DLT) term. Here $\hat{m}$ represents the unit magnetization vector, $\frac{d\hat{m}}{dt}$ is the time derivative of the $\hat{m}$; $\hat{H}$ is effective field consisting of external field ($H_{ext}$), magnetostatic field ($H_{demag}$), exchange field ($H_{exch}$), and anisotropic field ($H_{anis}$); r = $\frac{DLT}{DLT+FLT}$ is the parameter representing the relative strength between DLT and FLT. The dominance of DLT for the SOT magnetization switching in the FMS film is already verified in earlier investigations[13,29–31] and we used r = 0.9 in this investigation. $\hat{\sigma}$ is the vector representing spin polarization; $\gamma$ is gyromagnetic ratio; $\alpha$ is the damping constant; and J is the current density. $\zeta = \frac{\hbar}{2e}\frac{P}{M_s t}$ is the spin orbit torque coefficient, where $\hbar$ is the reduced plank constant, $e$ is the charge of the electron, $P$ is the spin polarization, $M_s$ is the saturation magnetization, and $t$ is the sample thickness.

We conducted micromagnetic simulations using the MuMax$^3$ package,[32] which allows systematic exploration of SOT behavior by adjusting parameters such as bias field direction, magnetic easy axis orientation, and carrier spin polarization. The simulations were performed on a 10 μm × 10 μm × 10 nm FMS film with a cell size of 5 nm × 5 nm × 2.5 nm. The simulation parameters, such as saturation magnetization ($M_s$) = 15×10$^3$ A/m, [33–35] exchange coupling constant ($A_{ex}$) = 0.06×10$^{-12}$ A/m$^2$, [34] anisotropy constant ($K_u$ = 1×10$^3$ J/m$^3$), [34] damping constant ($\alpha$ = 0.01) [36] and polarization factor $P_0 = 0.3$,[37] were chosen to model a typical (Ga,Mn)(As,P) FMS film with PMA.

**Symmetry breaking effects in the spin-orbit torque switching**

**(a) Effect of external bias field and its misalignment**

The most common method to break symmetry between "up" and "down" magnetization states in SOT switching for FM films with PMA is by applying an external bias field perpendicular to the carrier polarization within the film plane.[13] In an FMS film, carrier spin polarization depends on the crystal direction along which current flows, as shown in Fig. 1(a), where the Dresselhaus- and Rashba-type spin-orbit fields (SOFs) for the respective current directions (indicated by the black dashed arrow) are represented by red and blue arrows, respectively. A perpendicular configuration between bias field and spin polarization is achieved in (Ga,Mn)(As,P) films when current flows along the <110> direction with an in-plane bias field parallel to the current. The [1$\bar{1}$0] direction is defined as positive current ($+J_x$) and the bias field $B_x$ is positive (negative) for $+x$ ($-x$) direction as schematically shown in Fig. 1(b).

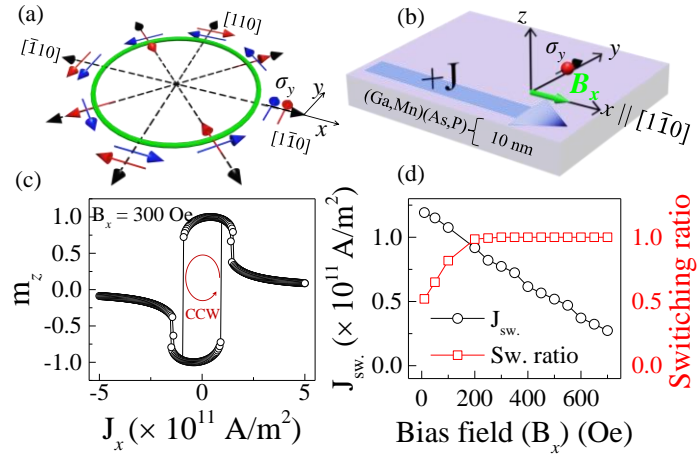

**Fig. 1** (a) Dresselhaus (red arrow) and Rashba (blue arrow) spin-orbit fields (SOFs) for the crystallographic directions along which the current (black dashed arrow) flows in a tensile-strained (Ga,Mn)(As,P) film. (b) Schematic of the simulated sample, illustrating the configuration for positive current along the $x$-axis (i.e., [1$\bar{1}$0] direction) and the corresponding polarization ($\sigma_y$) along the $y$-axis (i.e., [110] direction). (c) Current scan of the SOT hysteresis loop obtained under a +300 Oe external bias field. (d) Dependence of the SOT switching current density (black circles) and switching ratio (red squares) on the external bias field.

A SOT current scan hysteresis loop for this configuration was obtained by solving the LLG equation with a +300 Oe bias field, as shown in Fig. 1(c). The loop displays SOT switching with counterclockwise (CCW) chirality, consistent with prior reports for current scans along the [1$\bar{1}$0] direction under a +$B_x$.[13] The SOT switching chirality is changed clockwise (CW) when the sign of $B_x$ is reversed to negative (see supplementary section 1). The dependence of SOT switching on the bias field is shown in Fig. 1(d), where black circles and red squares present switching current density ($J_{sw.}$) and switching ratio (i.e., $m_z/m_{tot}$), respectively. The results show a monotonic decrease in switching current density and an increase in the switching ratio as the bias field strengthens, confirming the role of bias field in facilitating SOT magnetization switching. Note further that the SOT switching ratio reaches 1 above ~200 Oe (as shown by the red squares in Fig. 1(d)), indicating that a minimum bias field is required for complete SOT switching. The specific value of this threshold depends on material parameters, such as magnetization and magnetic anisotropy, in the FMS system.[38]

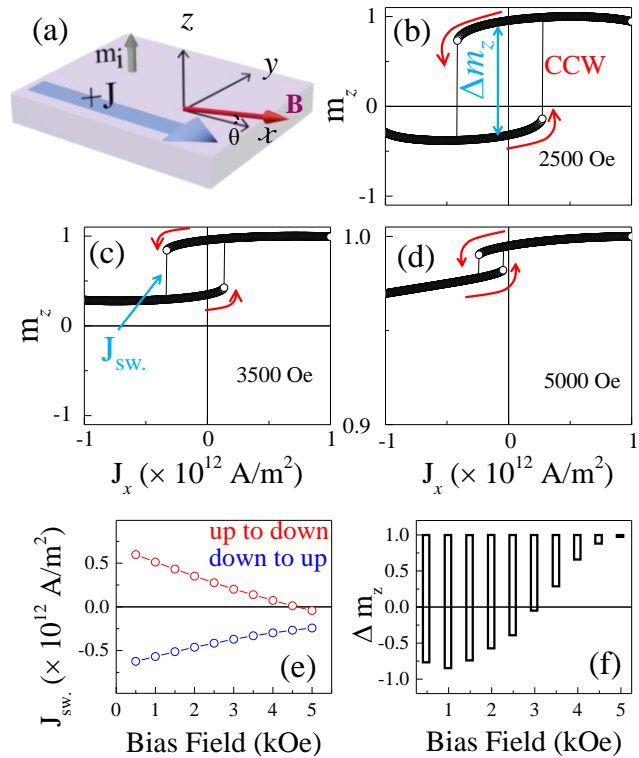

**Fig. 2.** (a) Schematic diagram illustrating the experimental configuration with a misalignment of the external bias field. The bias field is applied in the direction of the current (+$x$ direction) at an angle $\theta$ (~10°) relative to the film plane. (b-d) Current scan of the SOT magnetization switching hysteresis obtained at various external bias fields. Note the systematic changes in both switching current ($J_{sw.}$) and amplitude ($\Delta m_z$) with increasing external bias field. (e) Summary plot of the $J_{sw.}$ versus bias field for "up to down" switching (red open symbols) and "down to up" switching (blue open symbols). (f) Summary plot of SOT switching amplitude ($\Delta m_z$) versus bias field.

Even though the SOT switching behavior under in-plane bias field is well described above, slight misalignment of the bias field from the film plane is unavoidable in experiments. This misalignment introduces an out-of-plane bias field component that further breaks symmetry between the "up" and "down" magnetization states, affecting SOT switching behavior. Here we investigated how the out-of-plane component of bias field affects the SOT switching behavior in the FMS film with PMA. To simulate this experimental condition, we introduced a +10° tilt in the bias field direction from the film plane (see red arrow in Fig. 2(a)). Additionally, we allowed domain pinning energy fluctuations, representing variations in magnetic anisotropy, to model magnetization reversal through domain nucleation and expansion, which is known as a magnetization reversal process in FMS system.[39]

After initializing magnetization in the +z direction (white arrow marked with $m_i$ in Fig 2(a)), current scan hysteresis loops were obtained under various bias field strengths (Figs. 2(b)-(d)). As the bias field increased, the loops became asymmetrical, shifting leftward (toward negative current) and upward (toward positive magnetization). The switching current density for both "up-to-down" and "down-to-up" transitions became increasingly asymmetric as shown in Fig. 2(e). The vertical shift (upward) can be visualized from the summary plot of switching amplitude ($\Delta m_z$), which decreased with increasing bias field as shown in Fig. 2(f).

This behavior results from the positive out-of-plane component $+B_z$ of the bias field, which strengthens with increasing bias field and favors magnetization in the +z direction. Consequently, the transition from -z to +z requires less current, while the reverse transition is more difficult, leading to a leftward shift in the hysteresis loop as seen in Fig. 2(b)-(d). The upward shift, seen as a loop restricted to the positive half-plane above ~3 kOe (Fig. 2(f)), is caused by partial magnetization switching due to pinning field distribution.[39] As $+B_z$ increases, magnetization pinning toward the +z direction intensifies, reducing the portion of magnetization available for -z transitions and systematically shifting the loop upward.

To further investigate the impact of out-of-plane bias field components from misalignment, we performed simulations covering all combinations of bias field components ($\pm B_x$, $\pm B_z$) and magnetization initializations ($\pm z \parallel M_{ini}$). Schematic diagrams of these configurations are shown as insets in Fig. 3, where upward and downward misalignments are marked by red and blue arrows, respectively. The SOT switching hysteresis loops obtained under a 5 kOe bias field are plotted with red and blue symbols for upward and downward misalignments, respectively. The loops consistently shift in opposite directions, highlighting the critical role of bias field misalignment in SOT switching behavior. Note, however, that the shift direction of hysteresis is reversed when the sign of $B_x$ is reversed for the same initialization of magnetization (compare Fig. 3 (a) and (c) for $+z \parallel M_{ini}$ and Fig. 3 (b) and (d) for $-z \parallel M_{ini}$). This is because the $B_x$ primarily determines the switching chirality of the hysteresis loop. The $B_z$ component, meanwhile, assists or hinders the "up-to-down" and "down-to-up" transitions depending on its sign, while maintaining the chirality established by $B_x$. The opposite shift behaviors of hysteresis loops in Figs. 3(a) and 3(b) for $+B_x$, and Figs. 3(c) and 3(d) for $-B_x$, confirm this interplay. Furthermore, the shift behavior remains the same regardless of magnetization initialization direction, provided the sign of $B_x$ remains consistent (Figs. 3(a)-(b) for $+B_x$ and Figs. 3(c)-(d) for $-B_x$). This indicates that bias field misalignment effects are independent of the magnetization initialization in SOT switching.

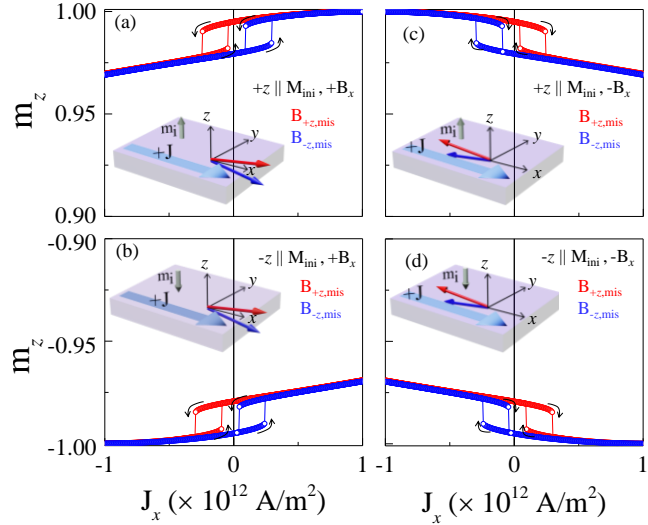

**Fig. 3**. Current scan of the SOT magnetization switching hysteresis obtained for various external bias field misalignments and initialization directions. (a) $+z$ initialization with $\pm B_z$ misalignment of the positive bias field. (b) $-z$ initialization with $\pm B_z$ misalignment of the positive bias field. (c) $+z$ initialization with $\pm B_z$ misalignment of the negative bias field. (d) $-z$ initialization with $\pm B_z$ misalignment of the negative bias field.

Recent observations of SOT switching hysteresis in FMS films under an in-plane bias field exhibited asymmetry in switching current between "up-to-down" and "down-to-up" transitions, along with an upward shift of the hysteresis loop. This asymmetry was attributed to a tilt in the magnetic easy axis from the vertical direction.[26] However, as demonstrated in Figs. 2 and 3, an out-of-plane bias field component caused by misalignment from the film plane can produce a similar asymmetric SOT hysteresis pattern. This highlights the need for careful interpretation of experimental data to accurately identify SOT behavior origins.

## (b) Interlayer coupling induced SOT switching

Symmetry breaking between the "up" and "down" magnetization states in FM films with PMA is often achieved via exchange coupling between two magnetic layers, which induces an internal bias field. In this study, we consider a (Ga,Mn)(As,P) film with a thin Mn-oxide-based magnetic layer formed through oxidation during thermal annealing.[25] The (Ga,Mn)(As,P) layer retains the material parameters described previously, while the Mn-oxide layer is modeled with an exchange coupling constant ($A_{ex}$) = $4\times10^{-12}$ A/m²[40] and an anisotropy constant ($K_u$ = $20\times10^3$ J/m³). The exchange coupling field acts similarly to an external bias field, influencing SOT switching when oriented perpendicular to the carrier polarization direction. As shown in Fig. 4(a), the coupling field from the Mn-oxide layer is aligned parallel to the [1$\bar{1}$0] current direction (+$x$ direction), realizing a perpendicular alignment with the carrier polarization direction (+$y$ direction) in the (Ga,Mn)(As,P) film. The interlayer exchange coupling between the two layers is represented by $\pm J_{coup}$, where the positive and negative signs indicate ferromagnetic (FM) and antiferromagnetic (AFM) coupling, respectively. In this simulation, the coupling strength is set to $|J_{coup}|$ = 10 x$10^{-15}$ J/m.

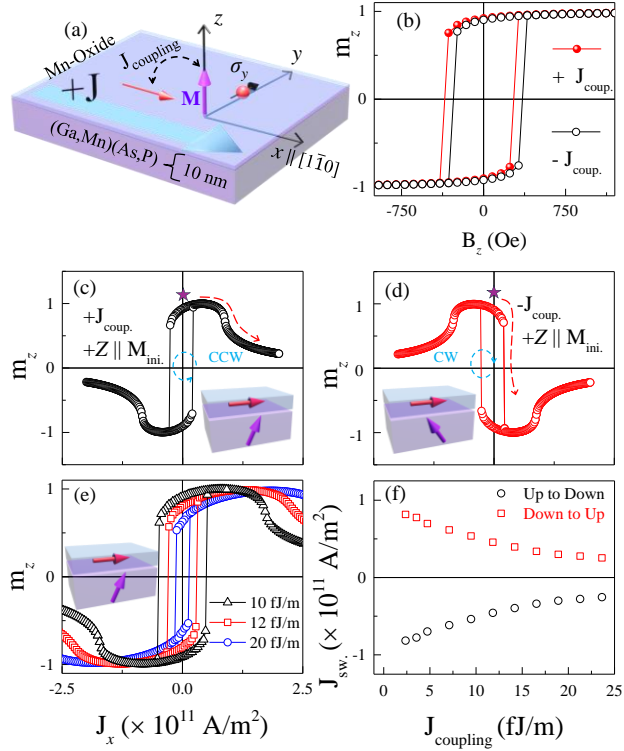

**Fig. 4.** (a) Schematic diagram of the (Ga,Mn)(As,P) film (10 nm) with a Mn-Oxide layer (2.5 nm) on top. The magnetizations of the Mn-Oxide (sky blue arrow) and (Ga,Mn)(As,P) (violet arrows) are initialized along the +$x$ and +$z$ directions, respectively. The two magnetizations are coupled by interlayer exchange coupling ($J_{coup.}$). (b) Field scan hysteresis loops of the Mn-Oxide/(Ga,Mn)(As,P) system for $J_{coup.}$ = 10 fJ/m. The hysteresis shifts in opposite directions

for opposite signs of $J_{coup}$. (c)-(d) Current scan of FF-SOT hysteresis loops obtained from the Mn-Oxide/(Ga,Mn)(As,P) exchange-coupled system with $\pm J_{coup}$ for two different configurations, as depicted in the insets of the panels. The initialized position is marked by a star symbol, and the first current scan direction is indicated by the red dashed arrow. (e) FF-SOT switching hysteresis obtained at three different interlayer coupling strengths for the $(+J_{coup}, +z \parallel M_{ini})$ configuration, shown in the schematic diagram in the inset. (f) Dependence of switching current density on the strength of the interlayer coupling. Red square and black circle symbols represent the switching current density for "up-to-down" and "down-to-up" switching, respectively.

To evaluate the impact of interlayer exchange coupling on magnetization reversal, we simulated field scan hysteresis loops, as shown in Fig. 4(b). The loops display opposite shifts depending on the type of coupling, indicated by the sign of $J_{coup}$, confirming an internal effective field induced by exchange coupling between the magnetic layers. From the hysteresis shift, we determined an effective coupling field of approximately 40 Oe, corresponding to $J_{coup} = 10$ fJ/m. Next, we investigated SOT switching behavior by solving the LLG equation under interlayer exchange coupling without an external bias field. After initializing the magnetization of the (Ga,Mn)(As,P) film in the $+z$ direction ($+z \parallel M_{ini}$), we obtained current scan hysteresis loops for both FM ($+J_{coup}$) and AFM ($-J_{coup}$) couplings as shown in Fig. 4(c)-(d). The results clearly show FF-SOT magnetization switching enabled by the effective field from interlayer exchange coupling between (Ga,Mn)(As,P) and Mn-oxide layers. Consistent with external bias field-assisted SOT switching, chirality is determined by the in-plane effective field direction. In the FM coupling case ($+J_{coup}$, field along $+x$), the switching chirality is counterclockwise (CCW) during current scans along the $x$-axis as seen in Fig. 4(c). Conversely, in the AFM coupling case ($-J_{coup}$, field along $-x$), switching chirality is clockwise (CW) (Fig. 4(d)). This demonstrates that the coupling type governs the in-plane effective field direction and, consequently, SOT switching chirality.

The FF-SOT switching achieved through interlayer exchange coupling offers distinct advantages over field-assisted (FA) SOT switching. First, the current density required for FF-SOT switching ($J = 0.5 \times 10^{11}$ A/m², Fig. 4(c)-(d)) is only half of that for field assisted (FA)-SOT switching ($J = 1 \times 10^{11}$ A/m², Fig. 1(c)). Second, while FA-SOT switching requires a minimum bias field of approximately 200 Oe for full magnetization reversal (Fig. 1(d)), FF-SOT switching only needs an effective exchange coupling field of about 40 Oe, five times smaller than the external bias field—highlighting the practical advantage of using interlayer exchange coupling in SOT devices.

To further examine SOT behavior, we varied the interlayer coupling strength. Fig. 4(e) presents SOT switching loops for different $+J_{coup}$ strengths from an up-initialized state, as depicted in the inset. A summary of the switching current density as a function of $+J_{coup}$ is shown in Fig. 4(f). Here, red square symbols indicate "down-to-up" switching in the positive current region, while black circle symbols represent "up-to-down" switching in the negative current region. The results reveal a systematic decrease in switching current density as coupling strength increases. This behavior mirrors the dependence of SOT switching current on the external bias field observed in FA-SOT switching (Fig. 1(d)), further confirming that interlayer exchange coupling plays a similar role to an external bias field in facilitating SOT switching in FMS films.

### (c) Role of out-of-plane component of polarization

As already mentioned, SOT from carriers with only in-plane spin polarization cannot reverse magnetization in FMS films with PMA. However, it is known that magnetization can be switched if the spin polarization has a component along the magnetic easy axis. [41,42] This implies that SOT switching in FMS films with PMA can be achieved when an out-of-plane spin polarization component is present.[43] To explore this behavior, we simulated SOT switching under the coexistence of out-of-plane ($z$-direction) and in-plane ($y$-direction) polarization components. In the simulation, we introduced an out-of-plane polarization component ($\sigma_z \hat{z}$) alongside the in-plane polarization ($\sigma_y \hat{y}$), resulting in a polarization vector ($\boldsymbol{\sigma}$) canted from the $y$-axis toward the $z$-axis. Fig. 5(a) illustrates this configuration, with the polarization direction represented by a black arrow with a red sphere. The polarization vector, with components along the $y$ and $z$ directions, is expressed as $\boldsymbol{\sigma} = (0, \sigma_y, \sigma_z)$, and its sign reverses when the current direction is reversed. The resulting SOT current scan hysteresis loop, shown in Fig. 5(b), was obtained using $\boldsymbol{\sigma} = (0, +\sigma_y, +\sigma_z)$ with a polarization ratio $\sigma_z/\sigma_y = 0.15$ for positive current (i.e., $+x$ direction) and without an external bias field. The loop clearly demonstrates FF-SOT switching, confirming the critical role of the out-of-plane polarization component ($\sigma_z$) in enabling field-free magnetization switching.

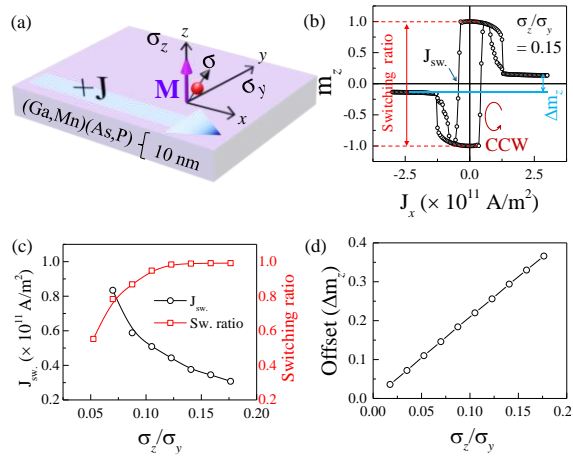

**Fig. 5.** (a) Schematic representation of the simulated sample. Spin polarization with $+y$ and $+z$ components are shown by a black arrow with a red sphere. The $\sigma_z$ component of polarization is parallel to the magnetization (**M**), represented by a thick violet arrow. (b) Current scan hysteresis loop obtained with a $\sigma_z/\sigma_y$ ratio of 0.15. The offset in the $m_z$ value, which appears at the opposite sign of the large current, is marked with a sky blue line and arrow. (c) Dependence of SOT switching current density ($J_{sw}$) and SOT switching ratio on the $\sigma_z/\sigma_y$ value, plotted with black and red open

symbols, respectively. (d) The offset monotonically increases with the strengthening of the $z$-component polarization (i.e., the $\sigma_z/\sigma_y$ ratio).

The observed FF-SOT switching behavior with $\sigma_z$ arises from the damping-like torque (DLT) contributed by $\sigma_z$, expressed as $(\hat{\mathbf{m}} \times \sigma_z \hat{\mathbf{z}} \times \hat{\mathbf{m}})$, which breaks the symmetry between the up and down magnetization states. This leads to transitions between these states, as shown in Fig. 5(b). For a positive current, the DLT from $+\sigma_z$ facilitates the "down-to-up" switching. When the current direction reverses, producing $-\sigma_z$, it assists the "up-to-down" transition. Consequently, FF-SOT switching occurs during the current scan. The chirality is counterclockwise (CCW) for $+\sigma_z$ under positive current (Fig. 5(b)) and becomes clockwise (CW) when $-\sigma_z$ is induced under the same current direction (see Supplementary Section 2). Thus, $\sigma_z$ solely determines the SOT switching chirality. Similar FF-SOT behavior induced by out-of-plane polarization has been experimentally observed in metal-based ferromagnetic systems. [44,45]

We further investigated how the strength of the $z$-component affects FF-SOT switching by varying the polarization ratio $\sigma_z/\sigma_y$ in the simulation. Hysteresis loops were obtained at different $\sigma_z/\sigma_y$ values (see Supplementary Section 3). The switching current density ($J_{sw}$) decreases as the $\sigma_z/\sigma_y$ ratio increases, indicating that higher $z$-polarization lowers the critical current. This trend is shown by black open symbols in Fig. 5(c), reinforcing the essential role of $\sigma_z$ in reducing $J_{sw}$. Additionally, the SOT switching ratio, plotted as red open circles in Fig. 5(c), increases with $\sigma_z/\sigma_y$ ratio, reaching ~100 % at $\sigma_z/\sigma_y \approx 0.125$. This indicates that the $z$-component of polarization must exceed a critical value to achieve complete SOT switching in FMS films.

Note that the non-zero $m_z$ component at large current, marked as an offset in Fig. 5(b), is a signature of FF-SOT switching induced by the additional $z$-component of spin polarization. At high current, where SOT dominates over anisotropy torque, magnetization of the film aligns along a direction set by two DLT components: $J(\hat{\mathbf{m}} \times \sigma_z \hat{\mathbf{z}} \times \hat{\mathbf{m}})$ and $J(\hat{\mathbf{m}} \times \sigma_y \hat{\mathbf{y}} \times \hat{\mathbf{m}})$. The $z$-component of DLT, $J(\hat{\mathbf{m}} \times \sigma_z \hat{\mathbf{z}} \times \hat{\mathbf{m}})$, causes the magnetization to tilt toward the $\pm z$-direction, depending on the polarization sign (i.e., $+\sigma_z$ for positive current and $-\sigma_z$ for negative current). Consequently, this produces a positive $m_z$ for positive current and a negative $m_z$ for negative current, creating the observed offset between opposite current directions. As shown in Fig. 5(d), this offset increases monotonically with the strength of the $z$-component polarization, represented by the $\sigma_z/\sigma_y$ ratio. This behavior further highlights the crucial role of out-of-plane polarization in FF-SOT switching.

### (d) SOT switching in FMS film with tilted vertical anisotropy

Magnetic anisotropy, which defines the magnetic easy axes of a FMS film, is influenced by various parameters, including chemical composition [46–48], strain, [49] or substrate type. [18,50] It can also be controlled through external methods.[51,52] The orientation of the magnetic easy axis is crucial for current-induced SOT magnetization control. For instance, while FF-SOT switching is achievable in FMS films with an in-plane magnetic easy axis, [42] it is not possible in those with a vertical easy axis.[13,29–31,53] This is because FF-SOT switching requires a collinear component of carrier spin polarization along the magnetic easy axis. In FMS films, where carrier spin polarization lies in the film plane,[41] it can have component along the magnetic easy axis if the anisotropy is in-plane. However, in FMS films with perpendicular magnetic anisotropy, the spin polarization is always orthogonal to the easy axis, preventing FF-SOT switching. When the magnetic easy axis tilts from the vertical direction, the spin polarization gains a collinear component along the easy axis, enabling FF-SOT switching even in FMS films with out-of-plane anisotropy. We investigated the SOT switching behavior in FMS films with tilted vertical easy axes as follows.

Consider the case shown in Fig. 6(a), where the $xy$-plane of the S($x$, $y$, $z$) coordinate system represents the film plane. The vertical magnetic easy axis tilts toward the carrier spin polarization direction ($\sigma_y$), indicated by the violet arrow, with the tilt angle $\theta$ measured from the $+z$ direction toward the $y$-direction. The polarization vector, generated by current along the $x$-axis, is defined as (0, $\sigma_y$, 0) within the film plane—representing the current along the [1$\bar{1}$0] crystallographic direction in the FMS film. SOT switching behavior was simulated for a tilt angle $\theta = 7°$ without an external bias field. The resulting current scan hysteresis loop, shown in Fig. 6(b), clearly demonstrates FF-SOT magnetization switching. This FF-SOT switching in the tilted-easy-axis FMS film occurs due to the spin-polarization component aligning with the magnetic easy axis, as discussed in the previous section.

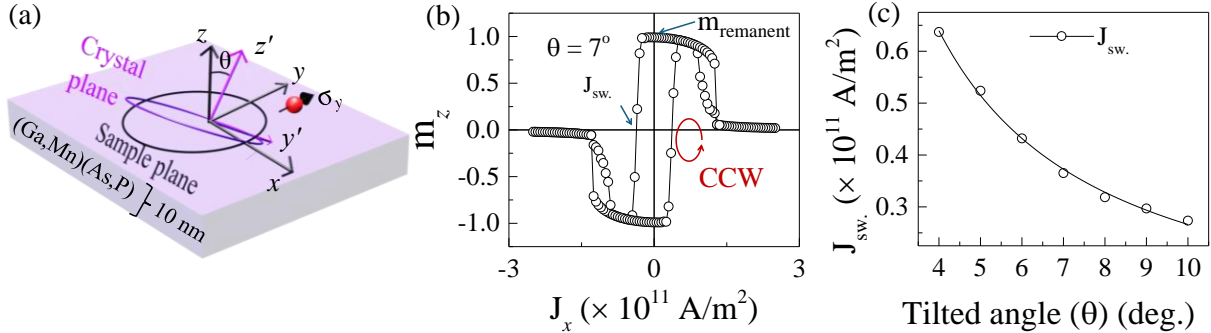

**Fig. 6**. (a) Schematic representation of the simulated sample with a tilted vertical anisotropy at an angle ($\theta$). The tilted anisotropy is represented as a tilted plane, indicated by a violet-colored frame, while the film plane is shown with a black-colored frame. The tilted angle of easy axis is denoted as $\theta$. (b) SOT current scan hysteresis loop for a tilted angle ($\theta = 7°$). The tilted plane provides a $+z'$ component of spin polarization ($\sigma$), which leads to the counterclockwise (CCW) chirality of the SOT current scan switching loop. (c) The switching current density ($J_{sw.}$) monotonically decreases as the tilted angle ($\theta$) increases.

To further analyze this behavior, we introduce an additional coordinate system S'($x'$, $y'$, $z'$), where the $z'$-axis aligns with the tilted magnetic easy axis (Fig. 6(a)). The tilt angle $\theta$ is defined as positive when directed toward +y and negative toward -y. The S(x, y, z) and S'(x', y', z') coordinate systems are related by a rotation around the x-axis. For a positive tilt angle $\theta$, the rotation transformation is expressed as:

$$R_x(\theta) = \begin{pmatrix} 1 & 0 & 0 \\ 0 & \cos\theta & -\sin\theta \\ 0 & \sin\theta & \cos\theta \end{pmatrix} \quad\quad\quad\quad (2).$$

Under such transformation, the polarization vector **σ** can be represented in the tilted coordinate system, S'(x', y', z') as:

$$\begin{pmatrix} \sigma'_x \\ \sigma'_y \\ \sigma'_z \end{pmatrix} = \begin{pmatrix} 1 & 0 & 0 \\ 0 & \cos\theta & -\sin\theta \\ 0 & \sin\theta & \cos\theta \end{pmatrix} \begin{pmatrix} 0 \\ \sigma_y \\ 0 \end{pmatrix} = \begin{pmatrix} 0 \\ \sigma_y \cos\theta \\ \sigma_y \sin\theta \end{pmatrix} \quad\quad (3)$$

In the tilted coordinate system S'(x', y', z'), the polarization vector σ has components $\sigma'_z = \sigma_y \sin\theta$ along the z'-axis (tilted easy axis direction) and $\sigma'_y = \sigma_y \cos\theta$ along the y'-axis. The $\sigma'_z$ component, aligned with the magnetic easy axis, drives FF-SOT magnetization switching as described previously. When the tilt angle θ is positive (toward +y), the polarization components are $\sigma'_z = \sigma_y \sin\theta$ and $\sigma'_y = \sigma_y \cos\theta$. This($+\sigma'_y$, $+\sigma'_z$) polarization combination under positive current produces FF-SOT switching with CCW chirality, as shown in Fig. 6(b). Conversely, for a negative tilt angle (toward -y), the components become $\sigma'_z = \sigma_y \sin(-\theta)$ (directed toward -z') and $\sigma'_y = \sigma_y \cos(-\theta)$ (toward +y'), forming a ($+\sigma'_y$, $-\sigma'_z$) polarization combination, which results in CW chirality (see Supplementary Section 4). This dependence of FF-SOT chirality on tilted anisotropy has been observed experimentally in metal-based magnetic systems.[22] The relationship between switching current density ($J_{sw.}$) and tilt angle is plotted in Fig. 6(c). $J_{sw.}$ decreases as the tilt angle increases, which is explained by the larger polarization component aligning with the magnetic easy axis as the tilt angle grows.

To investigate the influence of easy-axis tilt direction on SOT switching behavior, we considered three azimuthal angles ($\varphi$ = 90°, 45°, and 0°) with the same polar angle ($\theta$ = 7°), as depicted in Fig. 7(a)-(c), in which a bias field direction is shown by green arrows parallel to the current (i.e., x-direction). The resulting SOT current scan loops, with and without a bias field, are shown in Fig. 7(d)-(f). Two key features emerge from these simulations. First, FF-SOT switching occurs for $\varphi$ = 90° and $\varphi$ = 45°, but not for $\varphi$ = 0° (see black open symbols in Fig. 7(d)-(f)). Second, with a bias field (+$B_x$), the SOT loops are symmetric for $\varphi$ = 90° (Fig. 7(d)) but become asymmetric for $\varphi$ = 45° and $\varphi$ = 0° (Figs. 7(e)-(f)).

The presence of the z'-component of polarization enables FF-SOT switching at $\varphi$ = 90° and $\varphi$ = 45°, where current along the x-direction generates this component. However, at $\varphi$ = 0°, no z'-component is induced, preventing FF-SOT switching. Therefore, achieving FF-SOT switching in FMS films requires the magnetic easy axis to tilt toward the carrier polarization direction. The negative current shift of SOT loops at $\varphi$ = 45° and $\varphi$ = 0° under a bias field result from the z'- component of the bias field. This $z'$-component is strongest at $\varphi = 0°$, weakens as φ increases, and disappears at $\varphi = 90°$. Consequently, the asymmetry in the SOT loops systematically decreases as φ increases from 0° to 90°, where loops remain symmetric regardless of bias field strength.

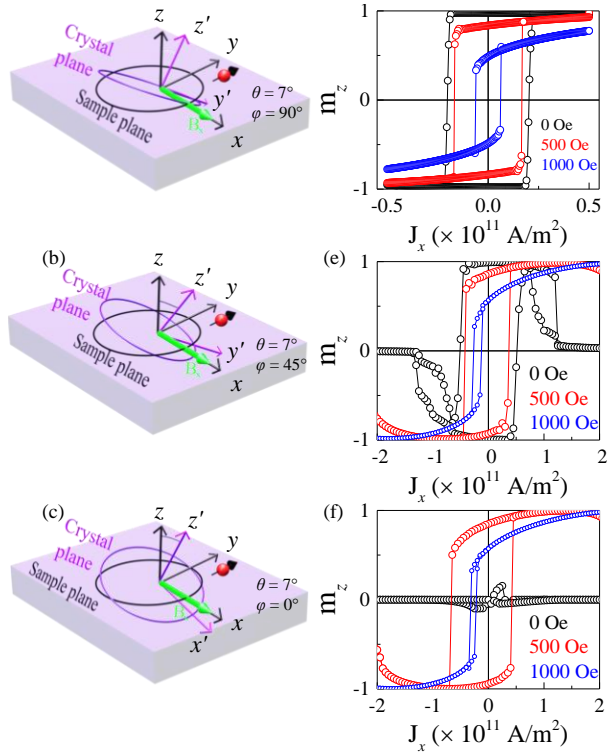

**Fig. 7.** Schematic configurations of the simulated sample with three different tilted vertical anisotropy directions: (a) $\varphi = 90°$, $\theta = 7°$, (b) $\varphi = 45°$, $\theta = 7°$, and (c) $\varphi = 0°$, $\theta = 7°$. The bias field (green thick arrow) is applied along the $+x$ direction. (d-f) Corresponding current scan loops in the presence of different bias field strengths for the three tilted configurations.

**Summary:**

This study explores symmetry-breaking effects in the SOT switching behavior of (Ga,Mn)(As,P) FMS films with PMA. The results highlight two key asymmetries introduced by the out-of-plane component of the bias field (caused by in-plane bias field misalignment): (1) increasing asymmetry in the current required for "up-to-down" and "down-to-up" switching with rising bias field magnitude, and (2) a shift of the SOT hysteresis toward positive or negative magnetization states, depending on the sign of the out-of-plane initialization. These findings emphasize the importance of carefully interpreting asymmetric hysteresis behavior in SOT experiments on FMS films. Simulations further reveal distinct impacts of internal symmetry-breaking mechanisms—including interlayer exchange coupling, out-of-plane spin polarization, and tilted magnetic anisotropy—on FF-SOT switching. For example, the chirality of SOT switching reverses depending on whether interlayer exchange coupling is ferromagnetic or antiferromagnetic. Additionally, in FMS films with out-of-plane polarization components, the chirality is determined solely by the sign of the out-of-plane component, regardless of the in-plane polarization. Similarly, FF-SOT switching in films with tilted vertical anisotropy is explained by considering the out-of-plane polarization component through a coordinate transformation toward the tilt direction. Here, the chirality of FF-SOT switching depends directly on the tilt direction of the easy axis: reversing the tilt direction reverses the switching chirality by changing the sign of the out-of-plane polarization component. These distinctive characteristics from different symmetry-breaking mechanisms can aid in identifying the cause of FF-SOT switching in experiments [22,44,45]. Furthermore, the findings indicate that tilting the vertical easy axis along the carrier polarization direction is essential for achieving FF-SOT switching—a critical consideration for FF-SOT experiments. This study reveals the crucial role of symmetry-breaking effects in FF-SOT switching and provides valuable insights for optimizing SOT efficiency and advancing spintronics technologies.

**Supplementary Material**

See Supplementary Material for discussion of the chirality of SOT current scan which depends on the external bias field direction, current direction, vertical magnetic anisotropy tiltation direction, additional z-component of polarization direction. Besides, we discussed the SOT magnetization switching for the case of exchange coupling direction aligns parallel to the polarization direction and also the case where the vertical magnetic anisotropy is titled perpendicular to the polarization direction.

**Acknowledgments**

This research was supported by Basic Science Research Program through the National Research Foundation of Korea (NRF) of Korea (2021R1A2C1003338); by the NRF under the BK21 FOUR program at Korea University, Initiative for Science Frontiers on Upcoming Challenges; by Korea University Grant; and by National Science Foundation Grant DMR 1905277.

**AUTHOR DECLARATIONS**
**Conflict of Interest**

The authors have no conflicts of interest to disclose.

## DATA AVAILABILITY

The data supporting the findings of this study are available from the corresponding author on request. Correspondence and requests for materials should be addressed to S.L. (email: slee3@korea.ac.kr).

**Symmetry-breaking effects on spin-orbit torque switching in ferromagnetic semiconductors with perpendicular magnetic anisotropy**

Apu Kumar Jana, Sanghoon Lee[*]

*Physics Department, Korea University, Seoul 136-701, Republic of Korea.*

**--: Supplementary:--**

**Supplementary section 1:**

We investigated the SOT switching chirality of the current scan hysteresis loops for different configurations of bias field and current directions. Fig. S1 shows current scans along the [1$\bar{1}$0] and the [$\bar{1}$10] direction (i.e., *x*-axis) under bias fields in the ±*x* direction. The switching chirality of the current scan loop alternates between counterclockwise (CCW) and clockwise (CW), depending on the sign of the bias field.

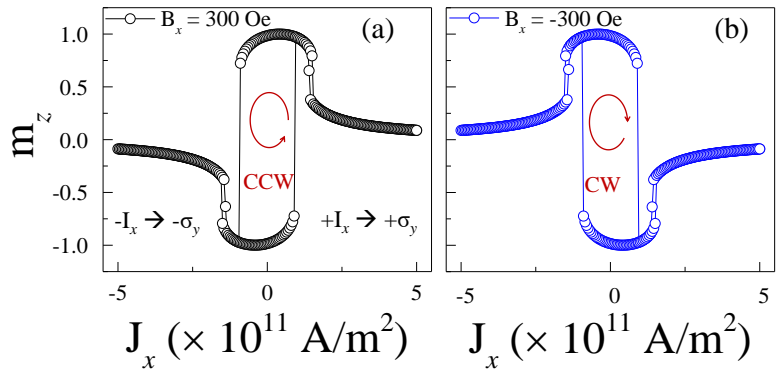

**Fig. S3.** SOT switching hysteresis loops for *x*-direction current scans under (a) positive and (b) negative bias fields.

**Supplementary section 2:**

We investigated the chirality of the SOT current scan loop in the presence of a ±z component of polarization, with the results shown in Fig. S2. When the *z*-component of polarization is **positive** for **positive current** (+*x* direction) and **negative** for **negative current** (-*x* direction), the SOT hysteresis loop exhibits **CCW switching chirality,** regardless of the in-plane polarization component ($\sigma_y$ or $\sigma_x$) (see Fig. S2(a) and (b)). Conversely, when the *z*-component of polarization is **negative** for **positive current** (+*x* direction) and **positive** for **negative current** (-*x* direction), the switching chirality **reverses to CW**, again independent of the in-plane polarization component (Fig. S2(c) and (d)). The **-z component of polarization** generates a **damping-like torque (DLT)** term $J(\hat{m} \times -\sigma_z \hat{z} \times \hat{m})$, initiating **"up to down"** switching, while the +*z* **component of polarization** results in **"down to up"** switching due to the DLT term $J(\hat{m} \times \sigma_z \hat{z} \times \hat{m})$. Thus, the SOT switching chirality is determined solely by the *z*-**component of polarization** ($\sigma_z$).

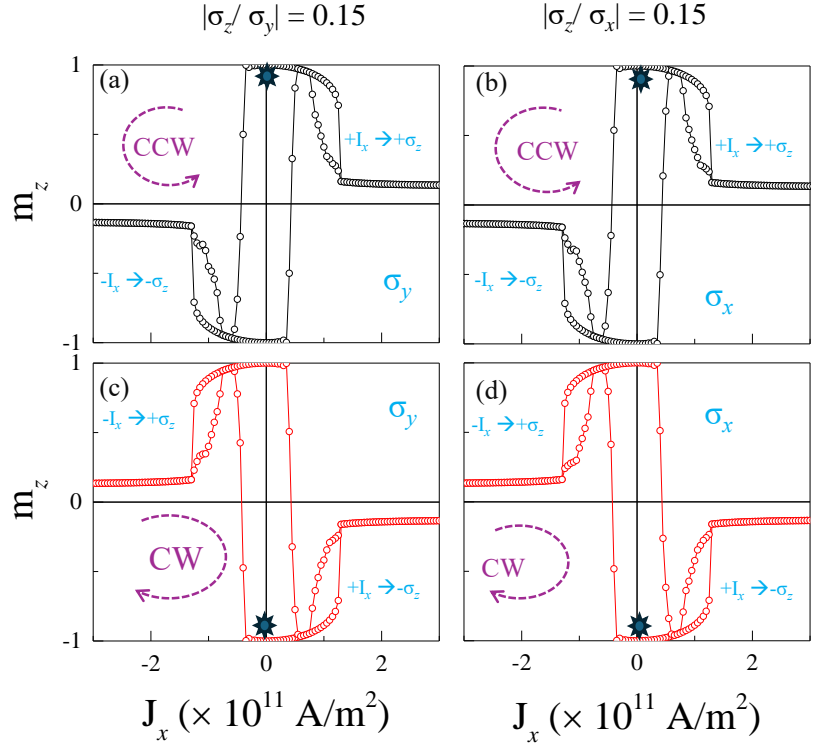

**Fig. S2.** FF-SOT current scan hysteresis with a **+z-component of polarization** for positive current, combined with in-plane polarization $\sigma_y$ (a) or $\sigma_x$ (b). FF-SOT current scan hysteresis with a **-z-component of polarization** for positive current, combined with in-plane polarization $\sigma_y$ (c) or $\sigma_x$ (d). Star symbols indicate the initial magnetization position.

**Supplementary section 3**

We investigated the FF-SOT current scan loop under varying strengths of the *z*-component polarization, as shown in Fig. S3. At large current values, the $m_z$ component remains nonzero, creating an offset between the two stabilized states at higher negative and positive currents, indicated by the dashed line. This offset increases **monotonically** with the strength of the *z*-component polarization, i.e., with a higher $\sigma_z/\sigma_y$ ratio. At higher $\sigma_z/\sigma_y$ ratios, the damping-like torque (DLT) term $(\hat{m} \times \sigma_z \hat{z} \times \hat{m})$ enhances the magnitude of $\pm m_z$ for positive and negative currents, respectively, resulting in a **larger offset ($\Delta m_z$)** between opposite current directions.

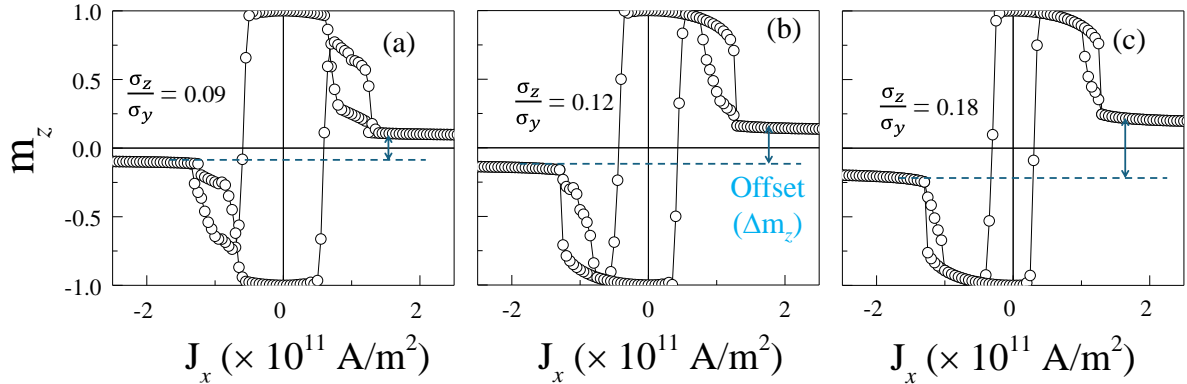

**Fig. S3.** SOT current scan hysteresis loops for (a) $\sigma_z/\sigma_y = 0.09$, (b) $\sigma_z/\sigma_y = 0.12$ and (c) $\sigma_z/\sigma_y = 0.18$ with $K_u = 5 \times 10^3$ J/m$^3$. The offset is indicated by the dashed line, showing an increase in $\Delta m_z$ as the $\sigma_z/\sigma_y$ ratio increases.

**Supplementary section 4:**

The presence of a vertical spin polarization component ($\sigma_z'$) along the magnetic easy axis drives FF-SOT magnetization switching, as described in the main text. Moreover, the direction of ($\sigma_z'$) determines the switching chirality in the current scan loop. When the tilt angle is **positive** (i.e., tilted toward $+\sigma_y$, as shown in Fig. S4(a)), the polarization component $\sigma'_z = \sigma_y \sin\theta$ induces **CCW switching chirality**, as seen in Fig. S4(b). Conversely, when the tilt angle is **negative** (i.e., tilted toward $-\sigma_y$, as shown in Fig. S4(c)), the polarization component $\sigma'_z = \sigma_y \sin(-\theta)$ points in the **-z' direction**, resulting in **CW switching chirality** in FF-SOT magnetization switching, as shown in Fig. S4(d).

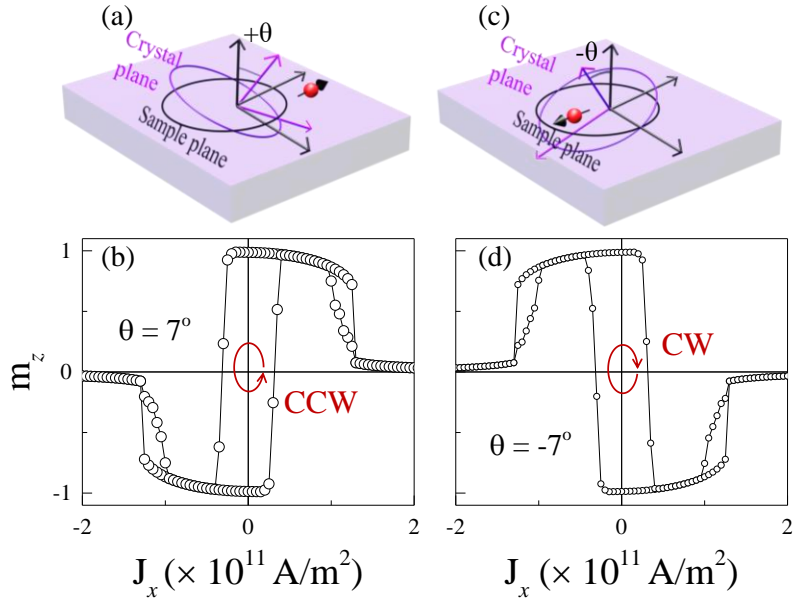

**Fig. S4.** Schematic diagram illustrating the tilting of the magnetic easy axis toward the **+y** (a) and **-y** (c) directions. **(b) and (d)** show SOT current scan hysteresis loops for tilting toward **+y** and **-y**, respectively, exhibiting **CCW** and **CW** SOT switching chirality.